\documentclass[prd,preprint,superscriptaddress,amsmath,amssymb,nofootinbib]{revtex4}
\usepackage{graphicx}
\usepackage{dcolumn}
\usepackage{bm}
\usepackage{amssymb}
\usepackage{amsmath}
\usepackage{epsfig}    
\usepackage{color}
\usepackage{slashed}
\usepackage{hhline}

\def\be{\begin{equation}}
\def\ee{\end{equation}}
\newcommand{\bea}{\begin{eqnarray}}
\newcommand{\eea}{\end{eqnarray}}

\newcommand{\lmlt}{L_\mu^{} - L_\tau^{}}

\begin{document}
\preprint{UME-PP-024}
\preprint{KYUSHU-HET-252}

\title{Multi muon/anti-muon signals via productions of gauge and scalar bosons 
in a $U(1)_{L_\mu - L_\tau}$ model at muonic colliders}

\author{Arindam Das}
\email{adas@particle.sci.hokudai.ac.jp}
\affiliation{Institute for the Advancement of Higher Education, Hokkaido University, Sapporo 060-0817, Japan}
\affiliation{Department of Physics, Hokkaido University, Sapporo 060-0810, Japan}

\author{Takaaki Nomura}
\email{nomura@scu.edu.cn}
\affiliation{College of Physics, Sichuan University, Chengdu 610065, China}

\author{Takashi Shimomura}
\email{shimomura@cc.miyazaki-u.ac.jp}
\affiliation{Faculty of Education, University of Miyazaki, 
1-1 Gakuen-Kibanadai-Nishi, Miyazaki 889-2192, Japan}
\affiliation{Department of Physics, Kyushu University, 744 Motooka, Nishi-ku, Fukuoka, 819-0395, Japan}

\date{\today}

\begin{abstract}
We discuss the discovery potential of a promising signals, $3 \times \mu^+ \mu^-$ at a $\mu^+ \mu^-$ collider 
and $\mu^+ \mu^+ \mu^+ \mu^+ \mu^- \mu^- $ at a $\mu^+ \mu^+$ collider,  
that are obtained via production of $Z'$ and a new scalar boson $\phi$ in a spontaneously broken local $U(1)_{L_\mu - L_\tau}$ model. We consider the $Z'$ associated production from the process $\mu^+ \mu^- \to \phi Z'$ in addition to a muonphilic $Z'$ fusion process $\mu^+ \mu^\pm \to \phi \mu^+ \mu^\pm$. The scalar boson is associated with $U(1)_{L_\mu - L_\tau}$ symmetry breaking and dominantly decays into $Z' Z'$ mode. We carry out numerical simulation analysis for signal and background processes to estimate a discovery significance for different benchmark points.
It is shown that our signal can be observed with integrated luminosity less than $\mathcal{O}(100)$ fb$^{-1}$ for both $\mu^+ \mu^-$ and $\mu^+ \mu^+$ colliders with more than 5-$\sigma$ significance.
\end{abstract}
\maketitle

\section{Introduction}

As an extension of the standard model(SM), introduction of new $U(1)$ gauge symmetry is one of the attractive possibilities.
New local $U(1)$ symmetry gives us an extra neutral gauge boson $Z'$ that can induce interesting phenomenology.
Among many possibilities of new $U(1)$ candidates, $U(1)_{L_\mu - L_\tau}$ symmetry is particularly interesting where 
$\mu$ and $\tau$ flavor leptons have charge $1$ and $-1$ under the symmetry~\cite{He:1990pn, He:1991qd}.
The $Z'$ boson associated with $U(1)_{L_\mu - L_\tau}$ couples to muon but not electron, and this property can be used to explain 
experimental anomalies related to muon.
When $Z'$ mass is light as $\mathcal{O}(10)$ to $\mathcal{O}(100)$ MeV, we can explain deviation of muon anomalous magnetic dipole moment (muon $g-2$) 
from the SM prediction~\cite{Bennett:2006fi, Abi:2021gix,Altmannshofer:2014pba,Lindner:2016bgg}.
Moreover we can apply this $Z'$ to explain lepton flavor non-universal anomalies observed in semi-leptonic $B$ meson decay associated with the process $b \to s \ell^+ \ell^-$~\cite{Hiller:2003js, Bobeth:2007dw, Aaij:2014ora,Aaij:2019wad, DescotesGenon:2012zf, Aaij:2015oid, Aaij:2013qta,Abdesselam:2016llu, Wehle:2016yoi,Aaij:2017vbb,Aaij:2021vac}
if $Z'$ is heavier than $\mathcal{O}(10)$ GeV~\cite{Altmannshofer:2014cfa, Crivellin:2015mga, Crivellin:2015lwa, Ko:2017yrd, Kumar:2020web,Han:2019diw,Chao:2021qxq,Baek:2019qte,Chen:2017usq,Borah:2021khc,Tuckler:2022fkz,Altmannshofer:2016jzy}~\footnote{ Recently the LHCb collaboration reported their observation regarding the lepton universality that is compatible with the SM prediction~\cite{LHCb:2022qnv,LHCb:2022zom} }.

Muonic colliders are quite suitable to explore $U(1)_{L_\mu - L_\tau}$ symmetry, 
which have been discussed with more increasing interests recently; $\mu^+ \mu^-$~\cite{MuonCollider:2022xlm,Aime:2022flm,Black:2022cth} as well as $\mu^+ \mu^+$~\cite{Hamada:2022mua, Hamada:2022uyn} colliders.
In fact, scattering processes at muon colliders mediated by $U(1)_{L_\mu - L_\tau}$ gauge boson are discussed exploring the testability of $Z'$ interactions~\cite{Huang:2021nkl}.
In addition to the scattering via $Z'$, we can also have productions of scalar boson  $\phi$ which is associated with $U(1)_{L_\mu - L_\tau}$ gauge symmetry breaking.
Such a scalar boson is inevitably induced when $U(1)_{L_\mu - L_\tau}$ symmetry is spontaneously broken by a vacuum expectation value (VEV) of some scalar field, 
and we have $\phi Z'Z'$ interaction.
The possible production processes are $\mu^+ \mu^- \to Z' \phi$ and muonphilic $Z'$ boson fusion of $\mu^+ \mu^- \to \mu^+ \mu^- Z'$ for $\mu^+ \mu^-$ collision, and muonphilic
$Z'$ fusion of $\mu^+ \mu^+ \to \mu^+ \mu^+ \phi$ for $\mu^+ \mu^+ $ collision~\footnote{The scalar boson productions in a $U(1)_{L_\mu - L_\tau}$ model at the LHC and the ILC are also discussed in refs.~\cite{Nomura:2020vnk,Nomura:2018yej}.  In addition, recently $Z'$ associated scalar production is discussed in the context of a $U(1)_{B_3-L_2}$ model~\cite{Allanach:2022blr}. }.
These new processes are perfectly suitable target for muonic colliders and we can explore 
evidence of $U(1)_{L_\mu - L_\tau}$ breaking mechanism by searching for signals from them.
In particular, we can get promising discovery potential for the parameter region that is motivated to explain $b \to s \ell^+ \ell^-$ anomalies; although recent LHCb observation regarding this lepton universality is compatible with the SM prediction this region is still interesting as the natural choice of gauge coupling scale $\mathcal{O}(0.1)$ to $\mathcal{O}(1)$.

In this paper, we investigate muonic signals that are obtained from scalar boson production processes with/without $Z'$ boson in a spontaneously broken $U(1)_{L_\mu - L_\tau}$ model.
Numerical simulations are carried out to estimate number of events for signal and background processes.
We then show discovery significance for benchmark points that are taken from the parameter region accommodating explanation of $b \to s \ell^+ \ell^-$ anomalies.

 This article is organized as follows.
In Sec. II, we discuss our model and show parameter region favored by explanation of $b \to s \ell^+ \ell^-$ anomalies.
In Sec. III, we carry out numerical calculation to estimate discovery potential of our signals.
 Finally we devote the summary of our findings.

\section{A spontaneously broken local $U(1)_{L_\mu - L_\tau}$ model}

\begin{table}[t]
  \begin{center}
    \begin{tabular}{|c|c|c|c|c|c|c|c|c|c||c|c|}\hline
      &
      $~~Q_L~~$ & 
      $~~u_R~~$ & 
      $~~d_R~~$ & 
      $~~L_e~~$ & 
      $~~L_\mu~~$ & 
      $~~L_\tau~~$ & 
      $~~e_R~~$ & $~~\mu_R~~$ & $~~\tau_R~~$ &
      $~~H~~$ & $~~\varphi~~$ \\ \hline
      $~~SU(2)_L~~$ & $\bf{2}$ & $\bf{1}$ & $\bf{1}$ & 
      $\bf{2}$ & $\bf{2}$ & $\bf{2}$ & $\bf{1}$ &$\bf{1}$ & $\bf{1}$ & $\bf{2}$ & $\bf{1}$ \\ \hline
      $~~U(1)_Y~~$ & $\frac{1}{6}$ & $\frac{2}{3}$ & $-\frac{1}{3}$ & 
      				$-\frac{1}{2}$ & $-\frac{1}{2}$ & $-\frac{1}{2}$ & $-1$ &$-1$ & $-1$ & $\frac{1}{2}$ & $0$ \\ \hline
      $~~U(1)_{L_\mu -L_\tau}~~$ & $0$ & $0$ & $0$ & $0$ & $1$ & $-1$ & $0$ &$1$ & $-1$ & $0$ & $1$ \\ \hline
    \end{tabular}
  \end{center}
  \caption{The gauge charge assignment of a local $U(1)_{L_\mu - L_\tau}$ model. }
  \label{tab:charge}
\end{table}

In this section we review a model with  $U(1)_{L_\mu - L_\tau}$ gauge symmetry which is spontaneously broken by a VEV of SM singlet scalar field $\varphi$.
We choose $U(1)_{L_\mu - L_\tau}$ charge of $\varphi$ as $1$ where the charge assignment is given in Table~\ref{tab:charge}.
In the table, $Q_L$, $u_R$ and $d_R$ are left-handed $SU(2)_L$ doublet quark, right-handed up-type quark and right-handed down-type quark that are the same as the SM ones. 
For lepton sector, $L_\alpha$ and $\alpha_R$ ($\alpha = e, \mu, \tau$) are left-handed lepton doublets and right-handed charged leptons which have flavor dependent $U(1)_{L_\mu - L_\tau}$ charges.
Scalar doublet $H$ is the SM Higgs field associated with electroweak symmetry breaking.

\subsection{Structure of the model}

The Lagrangian of our model is written by 
\begin{align}
\mathcal{L} =& \mathcal{L}_{\mathrm{SM}}- \frac{1}{4} Z'_{\mu \nu} Z'^{\mu \nu} 
+ g' Z'_\mu J^\mu_{\lmlt} - \frac{\epsilon}{2} B_{\mu \nu} Z'^{\mu \nu}  
+|D_\mu \varphi|^2 - V, \label{eq:Lagrangian}
\end{align}
where $\mathcal{L_{\mathrm{SM}}}$ denotes the SM Lagrangian except for Higgs potential, and the $L_\mu - L_\tau$ current $J^\mu_{\lmlt}$ and scalar potential are given as follows,
\begin{align}
J^\mu_{Z'} =& \bar L_\mu \gamma^\mu L_\mu + \bar \mu_R \gamma^\mu \mu_R - \bar L_\tau \gamma^\mu L_\tau - \bar \tau_R \gamma^\mu \tau_R, \label{eq:current} \\
V = &  -\mu_H^2 H^\dagger H - \mu_\varphi^2 \varphi^* \varphi + \frac{\lambda_H}{2} (H^\dagger H)^2 + \frac{\lambda_\varphi}{2} (\varphi^* \varphi)^2 + \lambda_{H \varphi} (H^\dagger H)(\varphi^* \varphi). \label{eq:scalar-potential}
\end{align}
The gauge fields and corresponding field strengths of $U(1)_{L_\mu - L_\tau}$ and $U(1)_Y$ are respectively denoted 
by $Z'$ and $B$. In Eq.\eqref{eq:Lagrangian}, we write the gauge coupling constant and the kinetic mixing parameter, respectively, by $g'$ and $\epsilon$.
In this paper, however, we ignore kinetic mixing parameter  due to the fact that  it is sufficiently small and does not affect phenomenology of our interest \cite{Nomura:2020vnk}.
In our scenario, the $U(1)_{L_\mu - L_\tau}$ gauge symmetry is spontaneously broken by VEV of $\varphi$.

The scalar fields $H$ and $\varphi$ are written by
\begin{equation}
H = \begin{pmatrix} G^+ \\ 
\frac{1}{\sqrt{2}} (v + \tilde h + i G^0) \end{pmatrix}, \quad \varphi = \frac{1}{\sqrt{2}} (v_\varphi + \tilde \phi + i G_{Z'}),
\label{eq:scalars}
\end{equation} 
where $G^+$, $G^0$ and $G_{Z'}$ are massless NG boson to be absorbed by gauge bosons $W^+$, $Z$ and $Z'$ while 
$\tilde{h}$ and $\tilde{\phi}$ represent the physical CP-even scalar bosons.
We obtain the VEVs of the scalar fields, $v$ and $v_\varphi$, from the stationary conditions 
$\partial V/ \partial v = \partial V/ \partial v_\varphi = 0$ such that
\begin{equation}
v = \sqrt{\frac{2 (\lambda_\varphi \mu_H^2 - \lambda_{H \varphi} \mu_\varphi^2 )}{\lambda_H \lambda_\varphi - \lambda_{H \varphi}^2 }}, \quad
v_\varphi = \sqrt{\frac{2 ( \lambda_H \mu_\varphi^2 - \lambda_{H \varphi} \mu_H^2) }{\lambda_H \lambda_\varphi - \lambda_{H \varphi}^2 }}. 
\end{equation}
Inserting Eq.\eqref{eq:scalars} into Eq.\eqref{eq:scalar-potential}, we find the squared mass terms for CP-even scalar bosons as
\begin{equation}
\mathcal{L} \supset  \frac{1}{2} \begin{pmatrix} \tilde h \\ \tilde \phi \end{pmatrix}^T \begin{pmatrix} \lambda_H v^2 & \lambda_{H \varphi} v v_\varphi \\  \lambda_{H \varphi} v v_\varphi  & \lambda_\varphi v_\varphi^2 \end{pmatrix} \begin{pmatrix} \tilde h \\ \tilde \phi \end{pmatrix}.
\end{equation} 
The squared mass matrix can be diagonalized by an orthogonal matrix and the mass eigenvalues are given by
\begin{equation}
m_{h, \phi}^2 = \frac{\lambda_H v^2 +\lambda_\varphi v_\varphi^2 }{4} \pm \frac{1}{4} \sqrt{\left( \lambda_H v^2 -\lambda_\varphi v_\varphi^2 \right)^2 + 4 \lambda_{H \varphi}^2 v^2 v_\varphi^2 }.
\end{equation}
We then find the corresponding mass eigenstates $h$ and $\phi$ as follows;   
\begin{equation}
\begin{pmatrix} h \\ \phi \end{pmatrix} = \begin{pmatrix} \cos \alpha & \sin \alpha \\ - \sin \alpha & \cos \alpha \end{pmatrix} \begin{pmatrix} \tilde h \\ \tilde \phi \end{pmatrix}, \quad
\tan 2 \alpha = \frac{2 \lambda_{H \varphi} v v_\varphi}{\lambda_H v^2 - \lambda_\varphi v_\varphi^2},
\end{equation}
where $\alpha$ is the mixing angle. In our analysis we assume $\sin \alpha \ll 1$, and $h$ is identified as the SM-like Higgs boson.

After $\varphi$ developing a VEV, $Z'$ boson gets its mass. 
Since we ignore kinetic mixing parameter, the mass is simply given by 
\begin{equation}
m_{Z'} = g' v_\varphi.
\end{equation}
We can also write interaction between $\phi$ and $Z'$ as 
\begin{equation}
\mathcal{L} \supset \frac12 m^2_{Z'} Z'^\mu Z'_\mu \left( 2 \frac{\phi}{v_\varphi} +  \frac{\phi^2}{v_\varphi^2} \right),
\end{equation}
where we adopted $\alpha \ll 1$.

\subsection{Parameter region favored by explanation of $b \to s \ell^+ \ell^-$ anomalies}

The $b \to s \ell^+ \ell^-$ anomalies can be explained when $Z'$ exchange induces effective interaction 
\begin{equation}
\mathcal{H}_{\rm eff} = - \frac{4 G_F}{\sqrt{2}} V_{tb} V^*_{ts} \frac{\alpha}{4 \pi} C_9^\mu (\bar b \gamma^\nu P_L s)(\bar \mu \gamma_\nu \mu),
\end{equation}
where $C_9^\mu$ is corresponding Wilson coefficient, $\alpha$ is the fine structure constant, $G_F$ is Fermi constant, and $V_{tb}$ and $V_{ts}$ are elements of CKM matrix. 
To induce the effective interaction, we need  $Z'^\mu \bar{b} \gamma_\mu P_L s$ coupling in addition to $Z'^\nu \bar \mu \gamma_\nu \mu$ one.
We can induce the former one by introducing new fields such as vector-like quarks(VLQs) and new scalar field through 
quark-VLQ mixing~\cite{Altmannshofer:2014cfa} or radiative corrections~\cite{Ko:2017yrd,Chen:2017usq}.
Here we assume the effective coupling 
\begin{equation}
\label{eq:effective_bs}
\mathcal{L}_{\rm eff} = g^L_{bs} Z'^\mu \overline{b_L} \gamma_\mu s_L,
\end{equation}
where $g^L_{bs}$ is considered to be free parameter.
We then obtain the Wilson coefficient from $Z'$ exchange, denoted by $C_9^{Z'}$, such that
\begin{equation}
\label{eq:C9Zp}
C_9^{Z'} \simeq - \frac{ \pi g' g^L_{bs}}{\sqrt{2} V_{tb} V^*_{ts} \alpha G_F m_{Z'_1}^2}. 
\end{equation}
The global fit in ref.~\cite{Alguero:2021anc} indicates $2 \sigma$ range of new physics contribution to $C_9^\mu$ as
\begin{equation}
-1.29 \leq \Delta C_9^\mu \leq -0.72,
\end{equation}
where the central value is $-1.01$.

\begin{figure}[t]
\begin{center}
\includegraphics[width=80mm]{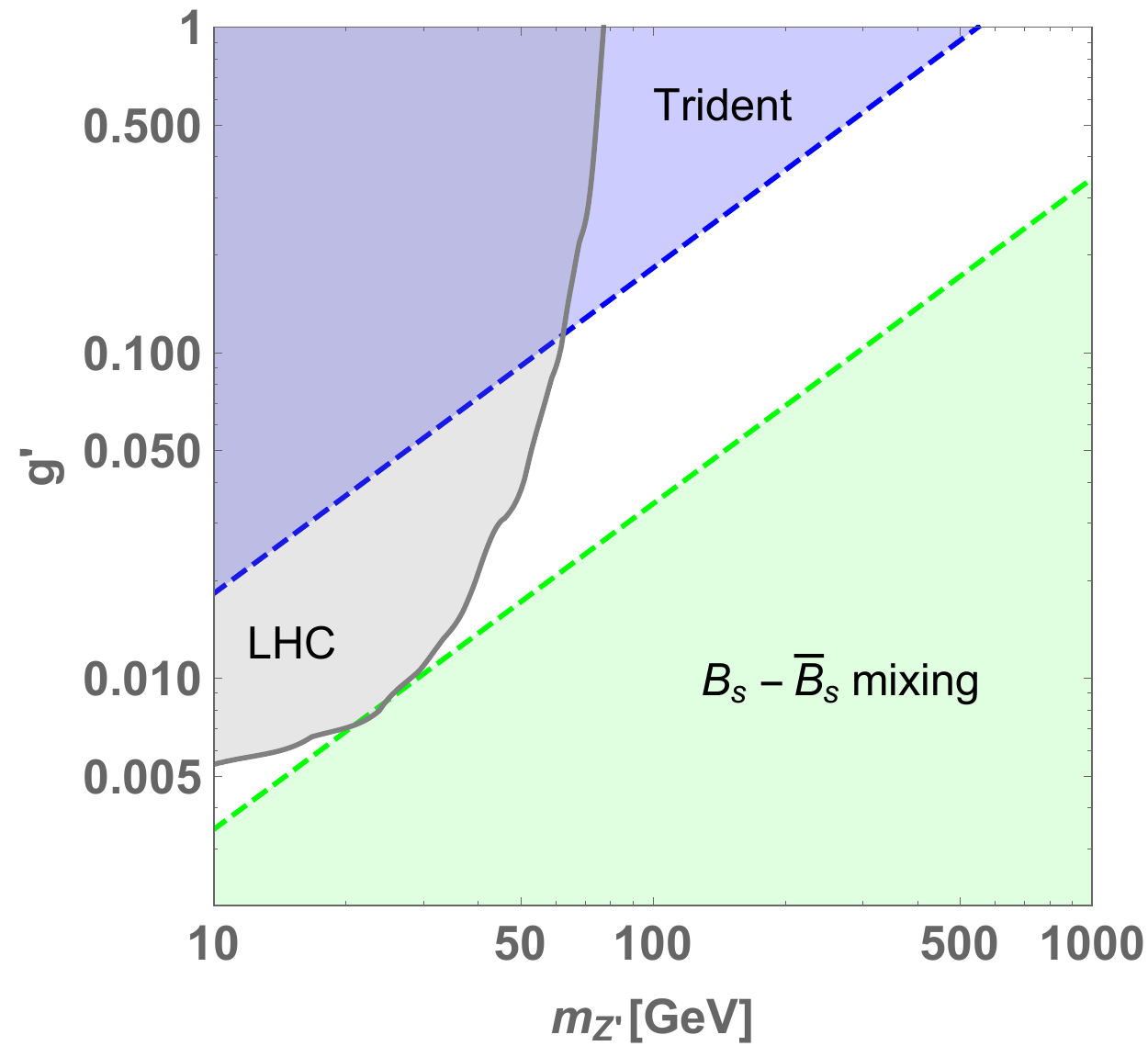}
\caption{Parameter region favored to explain $b \to s \ell^+ \ell^-$ anomalies. The gray and light blue region are excluded by the LHC data of $pp \to \mu^+ \mu^- Z'(\to \mu^+ \mu^-)$ search and neutrino trident respectively. The light green region is disfavored by constraint from $B_s$--$\overline{ B_s}$ mixing. } 
  \label{fig:gp-mzp}
\end{center}\end{figure}

The effective interactions in Eq.~\eqref{eq:effective_bs} also induce the mixing between $B_s$ and $\overline{B_s}$ mesons, 
and we should take a constraint from the mixing into account.
The ratio of $B_s$--$\overline{B_s}$ mixing between SM and $Z'$ contributions can be found as~\cite{Altmannshofer:2016jzy,Tuckler:2022fkz} 
\begin{equation}
\frac{M^{Z'}_{12}}{M^{\rm SM}_{12}} \simeq \frac{(g^L_{bs})^2}{m_{Z'}^2} \frac{16 \sqrt2  \pi^2}{g^2 G_F (V_{tb} V^*_{ts} )^2 S_0},
\end{equation} 
where $S_0 \simeq 2.3$ is the SM loop function~\cite{Inami:1980fz,Buchalla:1995vs}.
Then  $g^L_{bs}$ is rewritten in terms of $C_9^{Z'}$ in use of Eq.~\eqref{eq:C9Zp} such that
\begin{equation}
g^L_{bs} = -\sqrt{2} V_{tb} V^*_{ts} \frac{\alpha}{\pi} G_F C_9^{Z'} \frac{m_{Z'}^2}{g'}.
\end{equation}
Requiring $|M^{Z'}_{12}|/|M^{{\rm SM}}_{12}| < 0.12$~\cite{Charles:2020dfl}, we obtain the bound on $m_{Z'}$ as
\begin{equation}
m_{Z'} < 2.1 \ {\rm TeV} \times \frac{g'}{|C_9^{Z'}|}.
\end{equation}
In Fig.~\ref{fig:gp-mzp}, we show the parameter region favored to explain $b \to s \ell^+ \ell^-$ anomaly 
where some parameter region is excluded by the   
LHC constraint from $pp \to \mu^+ \mu^- Z'(\to \mu^+ \mu^-)$ process~\cite{CMS:2018yxg} and by CCFR constraints from neutrino trident production approximately given by $m_{Z'}/g' \gtrsim 550$ GeV~\cite{Altmannshofer:2014pba,CCFR:1991lpl}; if we consider $b \to s \ell^+ \ell^-$ process is consistent with the SM prediction we do not have constraint from green shaded region. Other constraints in the cointext of $L_{\mu}-L_{\tau}$ scenario can be found in \cite{Chun:2018ibr}. We will adopt parameter points on the favored region in our numerical simulation study.

\section{Signals at muonic colliders}

In this section we discuss discovery potential of signals via productions of the new scalar boson with/without $Z'$ boson in the model at muonic colliders.
As candidates of muonic colliders, we consider:
\begin{itemize}
\item $\mu^+ \mu^-$ collider with center of mass energy $\sqrt{s} = 3$ TeV~\cite{MuonCollider:2022xlm}.
\item $\mu^+ \mu^+$ collider with center of mass energy $\sqrt{s}= 2$ TeV~\cite{Hamada:2022mua}.
\end{itemize}
Here we focus on scalar boson production and the produced scalar boson $\phi$ dominantly decays into $Z'Z'$ mode. 
As a signal event, we consider the decay mode of $Z' \to \ell'^+ \ell'^- (\ell' = \mu, \tau)$ and choose muon event that gives us the clearest signals.
Therefore we consider signal events that includes $3 \times \mu^+ \mu^-$ and $\mu^+\mu^+\mu^+\mu^+ \mu^- \mu^-$ for $\mu^+ \mu^-$ and $\mu^+ \mu^+$ colliders~\footnote{We can also consider $Z'$ boson production such as pair(triple) production $\mu^+ \mu^- \to Z'Z'(Z'Z'Z')$ and muonphilic bremsstrahlung processes $\mu^+ \mu^\pm \to \mu^+ \mu^\pm + n Z'$. These signals can be distinguished from our signal, in principle, by considering mass reconstruction of the scalar boson, and we do not consider them. The full analysis including these $Z'$ productions will be done in our upcoming work(s). }.



\begin{figure}[t]
\begin{center}
\includegraphics[width=120mm]{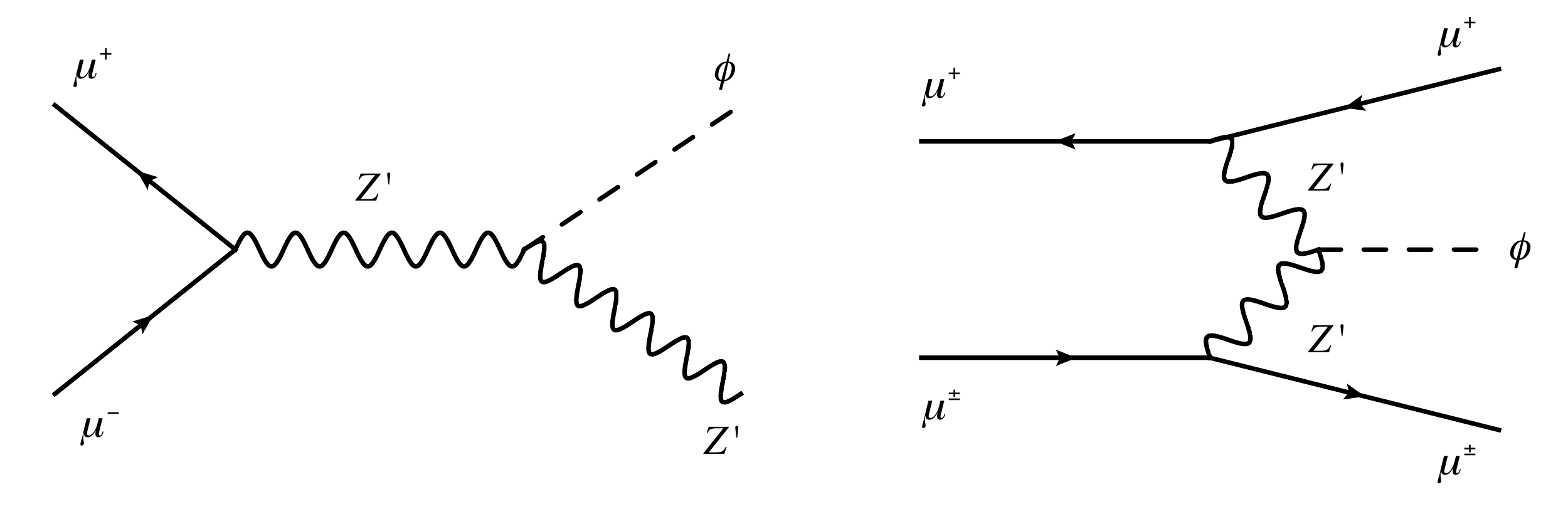}
\caption{Diagrams for the processes producing new scalar boson via $Z'$ interaction at $\mu^+\mu^-$ collider.}
  \label{fig:diagram}
\end{center}\end{figure}

\subsection{Scalar boson production processes}
\begin{figure}[t]
\begin{center}
\includegraphics[width=70mm]{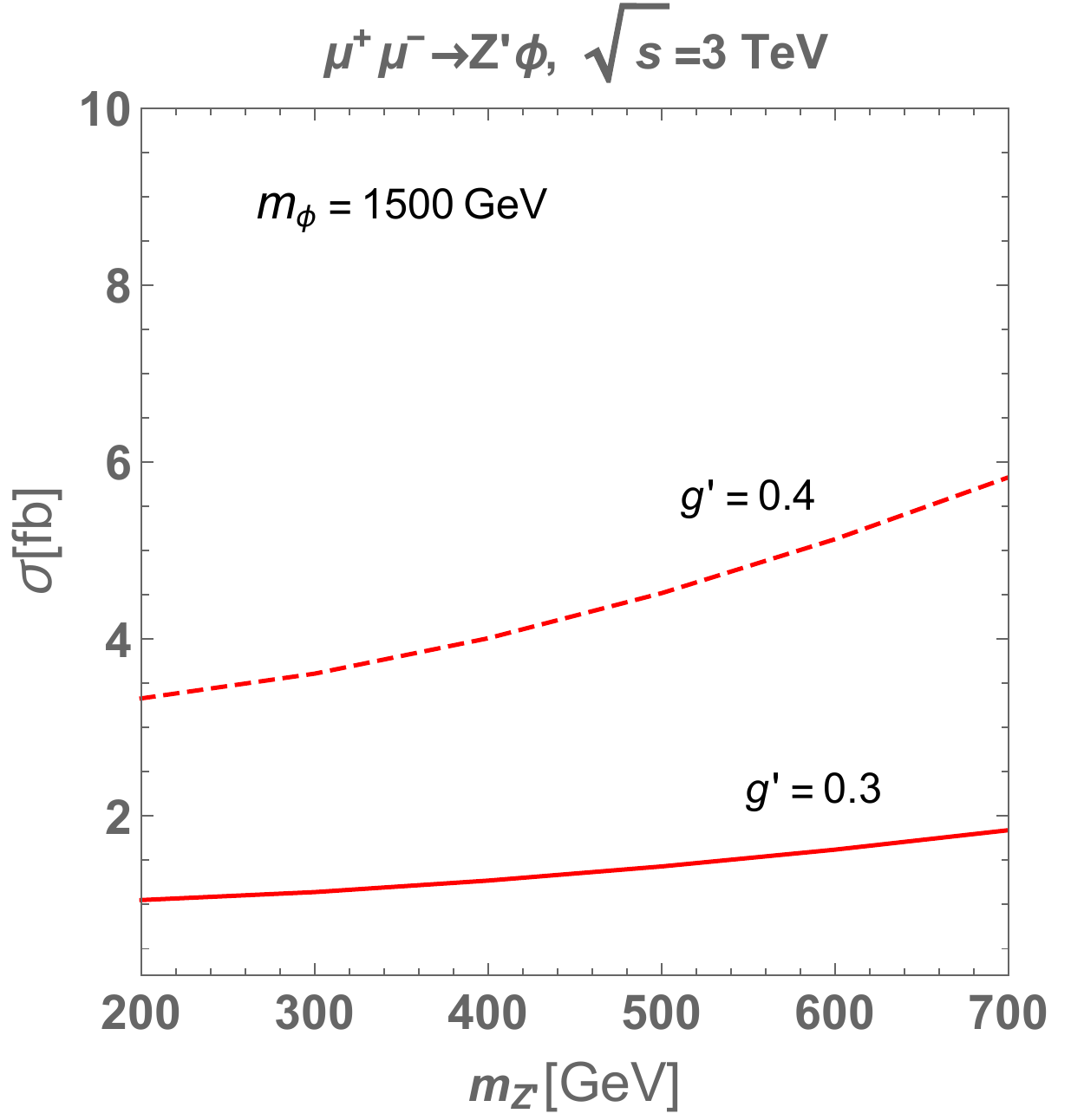} \quad
\includegraphics[width=70mm]{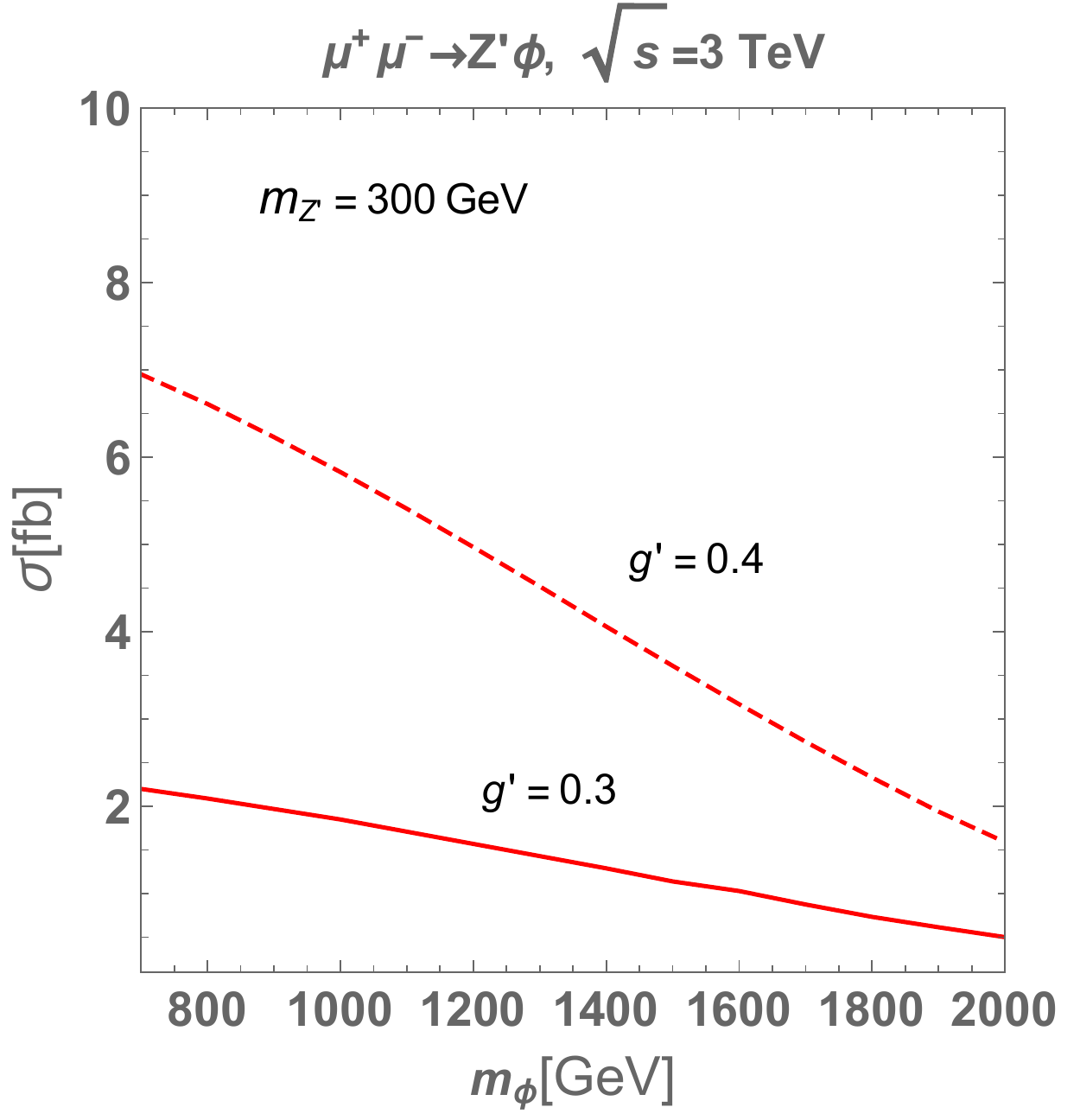}
\caption{The cross sections of $\mu^+ \mu^- \to \phi Z'$ process as a function of $m_\phi(m_{Z'})$ for left(right) plots where we apply $\sqrt{s} = 3$ TeV.} 
  \label{fig:CX1}
\end{center}\end{figure}
\begin{figure}[t]
\begin{center}
\includegraphics[width=70mm]{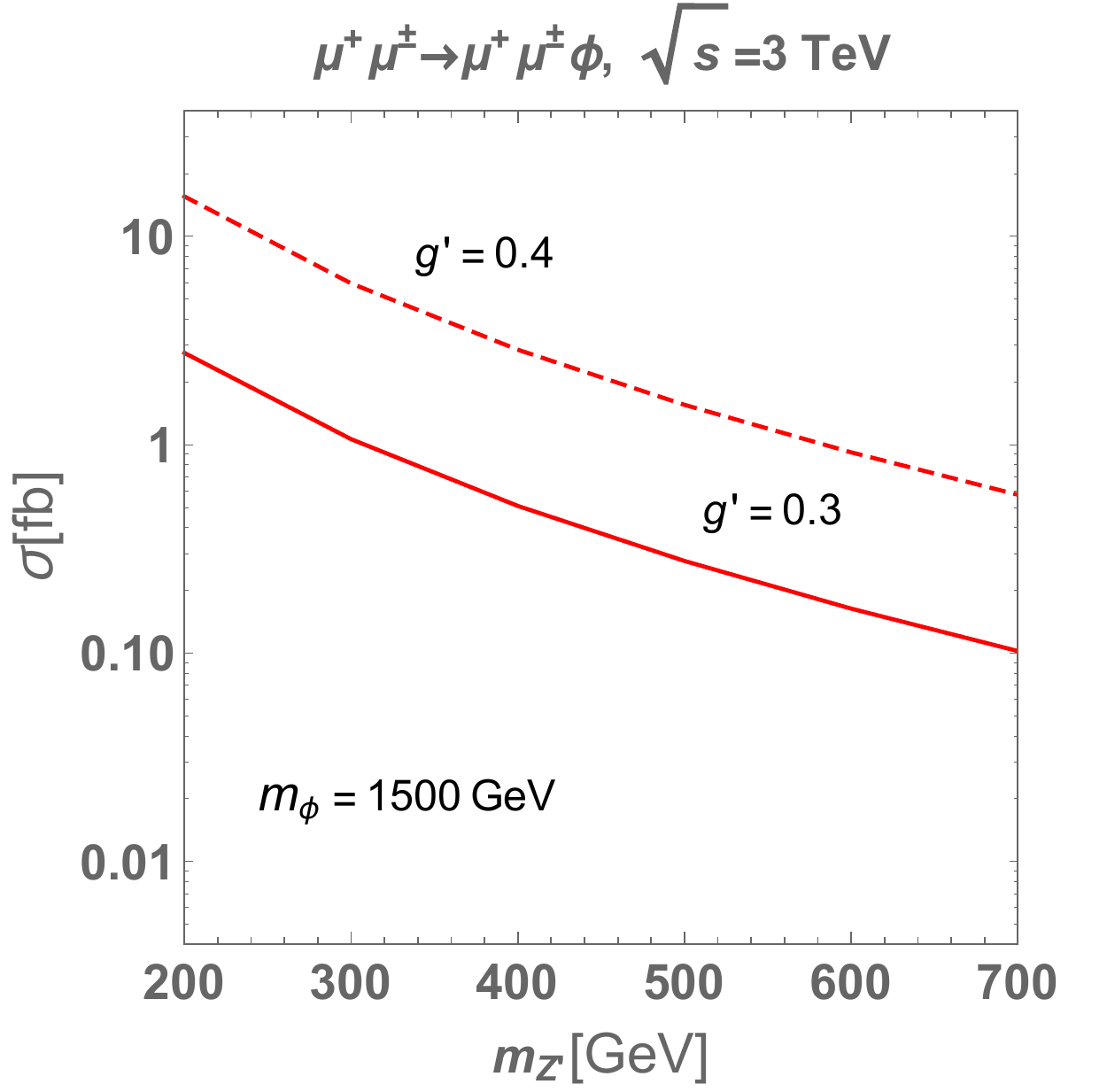} \quad
\includegraphics[width=70mm]{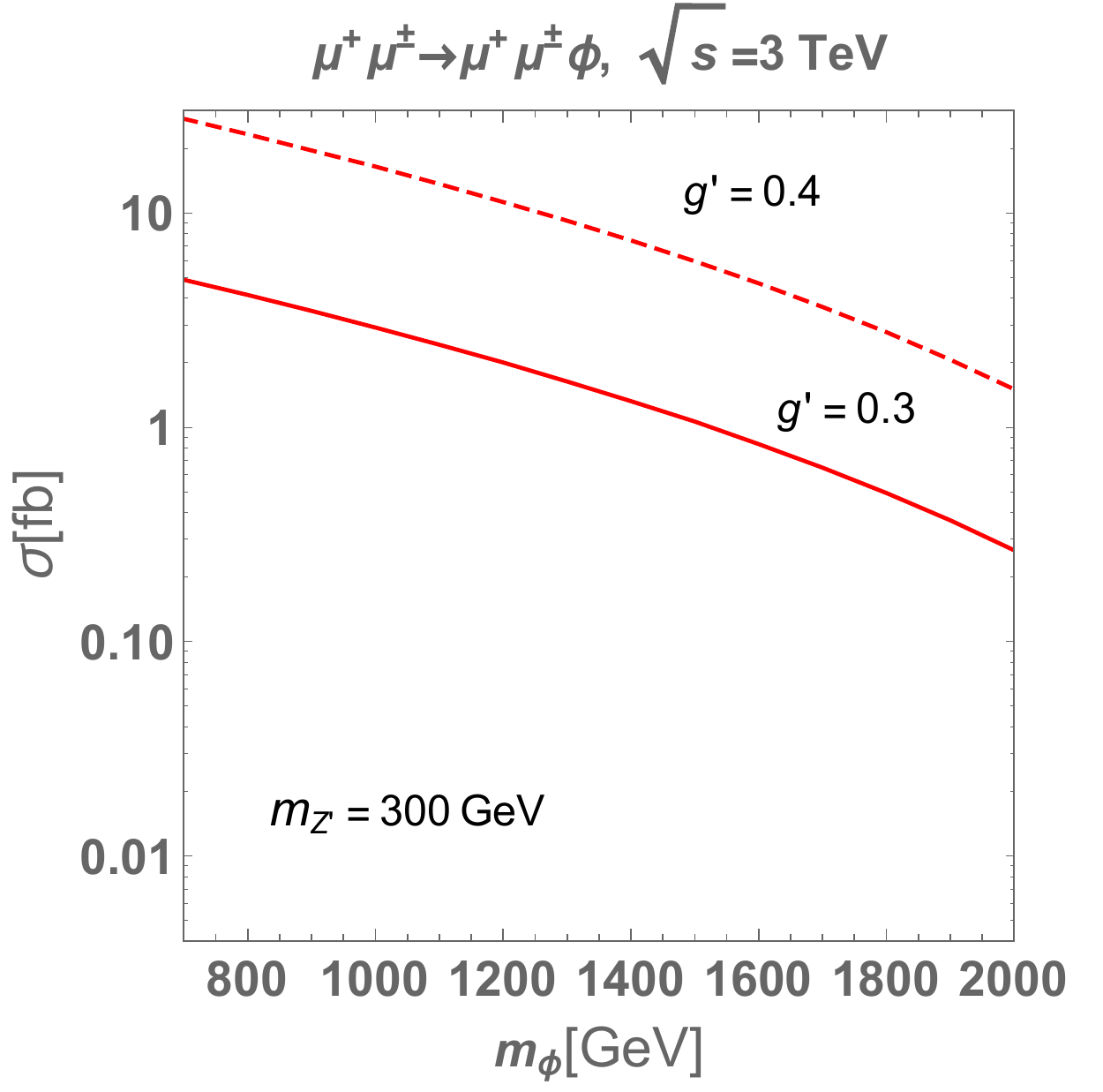}
\caption{The cross sections of $Z'$ fusion process $\mu^+ \mu^\pm \to \mu^+ \mu^\pm \phi$ as a function of $m_\phi(m_{Z'})$ for left(right) plots where we apply $\sqrt{s} = 3$ TeV.} 
  \label{fig:CX2}
\end{center}\end{figure}
\begin{figure}[t]
\begin{center}
\includegraphics[width=70mm]{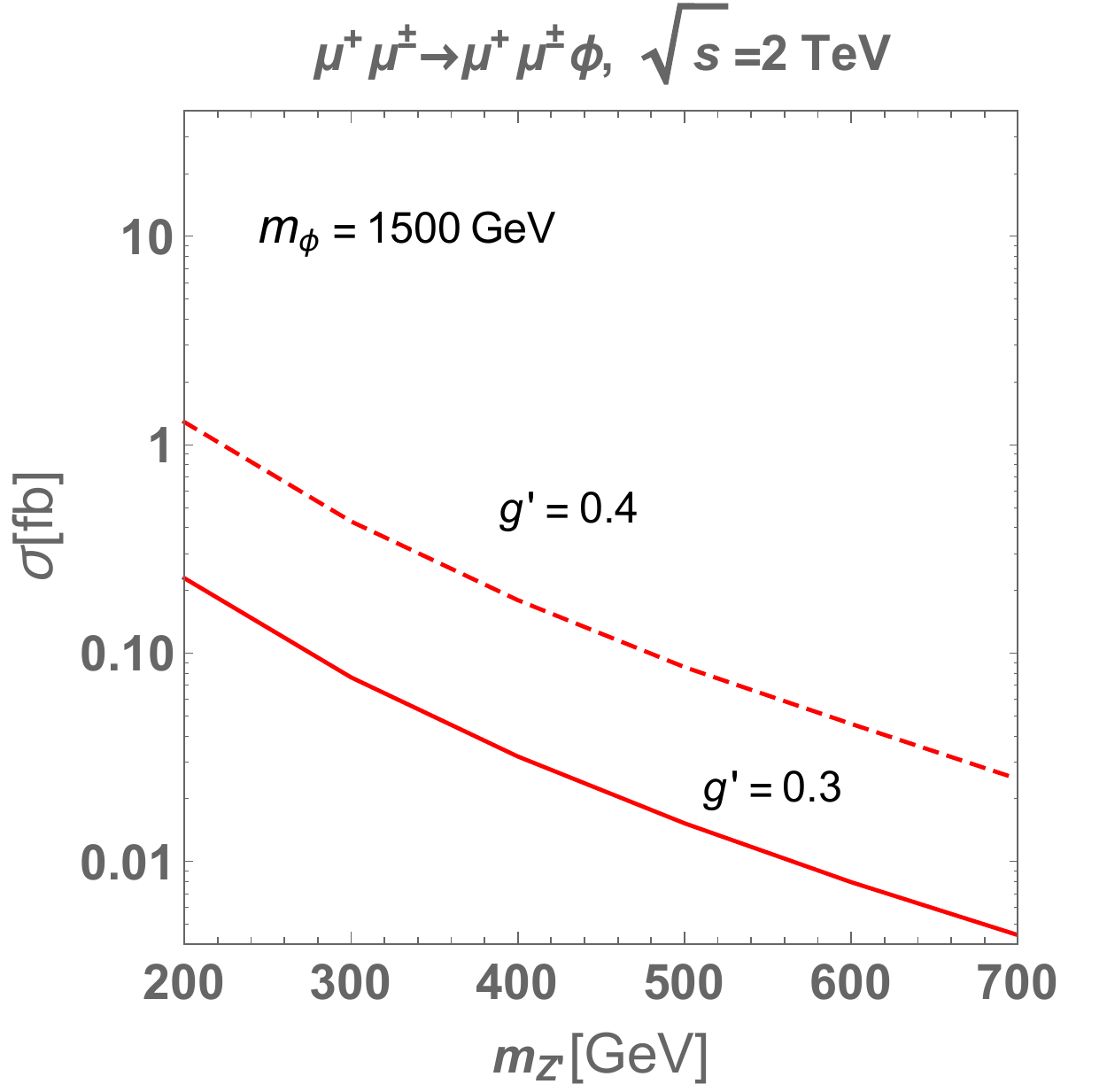} \quad
\includegraphics[width=70mm]{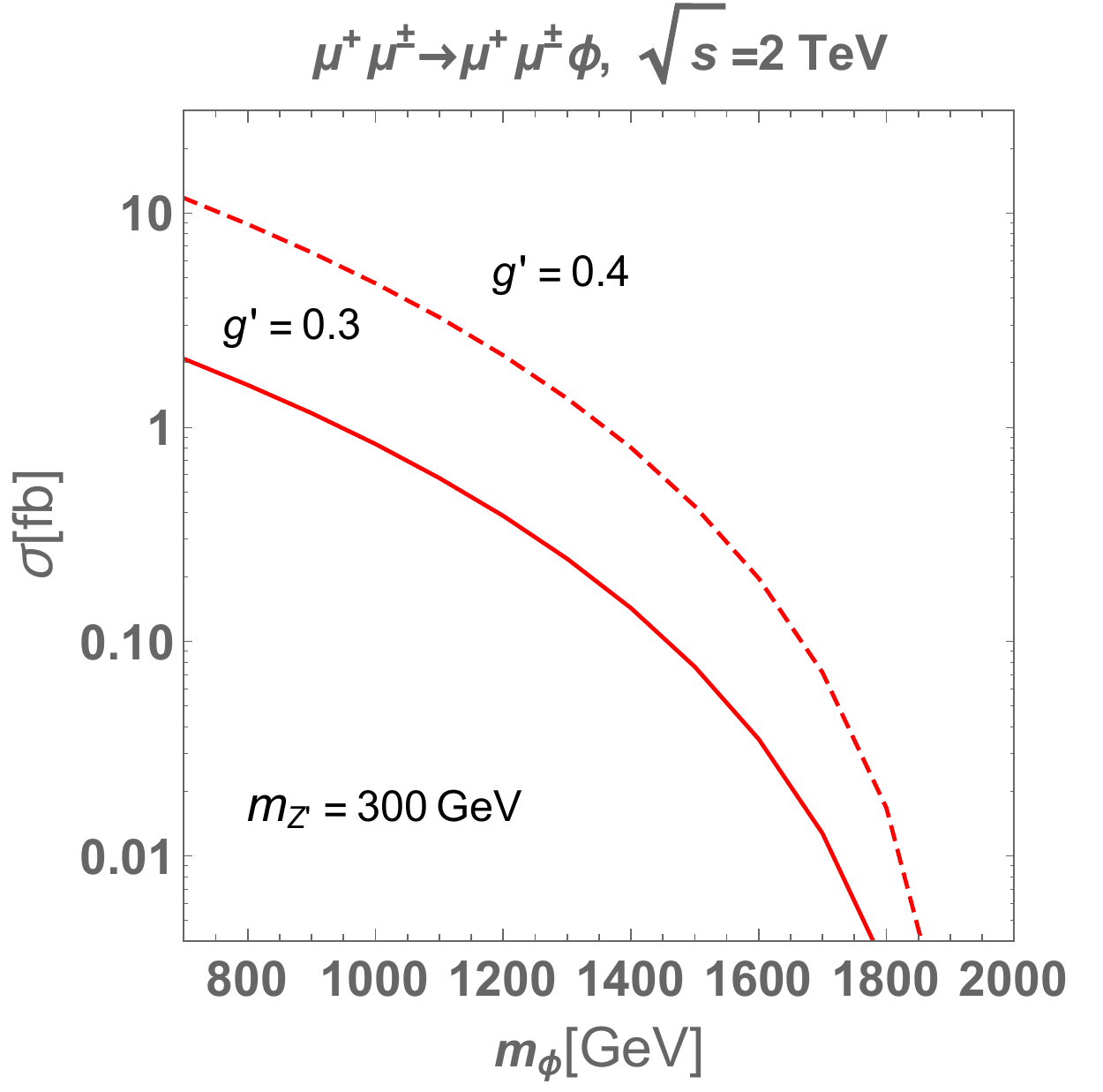}
\caption{The cross sections of $Z'$ fusion process $\mu^+ \mu^\pm \to \mu^+ \mu^\pm \phi$ as a function of $m_\phi(m_{Z'})$ for left(right) plots where we apply $\sqrt{s} = 2$ TeV.} 
  \label{fig:CX3}
\end{center}\end{figure}

The scalar boson $\phi$ can be produced by $\mu^+ \mu^-$ collision through the processes given in Fig.~\ref{fig:diagram}. 
The left diagram in the figure is $Z'$ associated production while the right diagram shows $Z'$ boson fusion production.
On the other hand the scalar boson can be produced only through $Z'$ fusion in the $\mu^+ \mu^+$ collision case.

We estimate the production cross sections for each process using {\it CalcHEP 3.7}~\cite{Belyaev:2012qa} implementing relevant interactions. 
The $Z'$ associated production cross sections are shown in Fig.~\ref{fig:CX1} for $\sqrt{s} = 3$ TeV as functions of $m_{\phi} (m_{Z'})$ in left(right) panel where we fixed other parameters as indicated on the plots.
Note that the cross section increases when $m_{Z'}$ becomes heavy for fixed $m_{\phi}$ since $Z'Z' \phi$ coupling is proportional to $m_{Z'}$.
The production cross sections of $Z'$ fusion process are also shown in Fig.~\ref{fig:CX2} for $\sqrt{s} =3$ TeV as functions of $m_{\phi} (m_{Z'})$ for left(right) panel where we fixed other parameters as indicated on the plots.
In this case the cross section decreases for heavier $Z'$ since we have two $Z'$ propagators in the diagram.
In addition, in Fig.~\ref{fig:CX3}, we show cross section of $Z'$ fusion for $\sqrt{s} = 2$ TeV with the same parameters as Fig.~\ref{fig:CX2}.

\subsection{Numerical simulations}

Firstly we discuss our signals from $\mu^+ \mu^-$ and $\mu^+ \mu^+$ collisions and corresponding background (BG) processes to estimate the discovery potential.

(1) $\mu^+ \mu^-$ collision case:
The signal and BG processes in this case are 
\begin{align}
& \text{Signal processes:} \quad  \mu^+ \mu^- \to Z' \phi, \quad \mu^+ \mu^- \to \mu^+ \mu^- \phi, \\
& \text{BG processes:} \quad \mu^+ \mu^- \to \mu^+ \mu^- Z Z, \ W^+ W^- Z Z, \ Z Z Z Z, 
\end{align}
where the BG processes provide $3 \times \mu^+ \mu^-$ events after decays of $W^\pm/Z$ bosons.
For signal event we include $Z'$ fusion process and $\mu^+ \mu^- \to Z'^{(*)}(\to \ell'^+ \ell'^-) \phi$ including both off-shell and on-shell $Z'$.
For the BG processes we obtain following cross sections at $\sqrt{s} = 3$ TeV estimated by {\tt MADGRAPH5}~\cite{Alwall:2014hca};
\begin{equation}
\sigma(\mu^+ \mu^- ZZ) = 3.12 \ {\rm fb}, \quad \sigma(W^+ W^- ZZ) = 1.20 \ {\rm fb}, \quad \sigma(ZZZZ) = 3.11 \times 10^{-3} \ {\rm fb}.
\end{equation}
The dominant BG is $\mu^+ \mu^- ZZ$ mode since we obtain the cross section of the second one with muonic $W^\pm$ decay as $\sigma(W^+ W^- ZZ) BR(W^- \to \mu^- \bar \nu_\mu) BR(W^+ \to \mu^+ \nu_\mu) \sim 0.013$ fb.
Then scalar boson $\phi$ decays into $Z'Z'$ mode with almost 100$\%$ BR when this mode is kinematically allowed.
The new gauge boson $Z'$ decays into $\mu^+ \mu^-$, $\tau^+ \tau^-$ and $\{\nu_\mu \bar \nu_\mu, \nu_\tau \bar \nu_\tau\}$ modes where each mode has BR$ = 0.33$, assuming neutrinos are Majorana particles and ignoring kinetic mixing effect.
Thus our signal event is $3 \times \mu^+ \mu^-$.

(2) $\mu^+ \mu^+$ collision case:
The signal and BG processes in this case are 
\begin{align}
& \text{Signal process:} \quad \mu^+ \mu^+ \to \mu^+ \mu^+ \phi, \\
& \text{BG process: } \quad \mu^+ \mu^+ \to \mu^+ \mu^+ Z Z,
\end{align}
where cross section of BG is 0.32 fb at $\sqrt{s} = 2$ TeV estimated by {\tt MADGRAPH5}.
The following decays of $\phi$ and $Z'$ are the same as case (1) inducing $\mu^+ \mu^+ \mu^+ \mu^+ \mu^- \mu^-$ signal event.

We next generate events for signal and BG using {\tt MADGRAPH5} implementing the model in use of FeynRules 2.0 \cite{Alloul:2013bka}.
In addition the {\tt PYTHIA\,8}~\cite{Sjostrand:2014zea}  is applied to deal with hadronization effects,  the  initial-state radiation (ISR) and final-state radiation (FSR) effects and the decays of SM particles, 
and we apply {\tt Delphes}~\cite{deFavereau:2013fsa} detector level simulation.
At the detector level, we select events as
\begin{align}
& 3 \times \mu^+ \mu^- \ : \ \text{for $\mu^+ \mu^-$ collider case}, \nonumber \\
&  \mu^+\mu^+\mu^+\mu^+ \mu^-\mu^- \ : \ \text{for $\mu^+ \mu^+$ collider case}.
\end{align} 
For generated events, we also impose following basic kinematical cuts
\begin{align}
\label{eq:Selection1}
 p_T(\mu^\pm) > 10 \ {\rm GeV}, \quad \eta(\mu^\pm) < 2.5 \, ,
\end{align} 
where $\eta = 1/2 \ln (\tan \theta/2)$ is the pseudo-rapidity with $\theta$ being the scattering angle in the laboratory frame and $p_T$ denotes transverse momentum.
As we see below, basic selection reduces BG event sufficiently and we do not discuss further details of kinematic distribution.
The information of kinematic distribution will be useful to identify our new particle $Z'$ and $\phi$, but it is beyond the scope of this work and 
we focus on discovery potential.

\begin{center} 
\begin{table}[t]
\begin{tabular}{|c|c|c|}\hline
\multicolumn{3}{|c|}{$\mu^+ \mu^-$ collision ($\sqrt{s}=3$ TeV)} \\ \hline
  & signal & BG   \\ \hline 
$N_{\rm ev}$ (without selection) &  205 & $433$   \\ \hline
$N_{\rm ev}$ (with selection) & 8.02 & 0.00827   \\ \hline
$S$ & \multicolumn{2}{|c|}{9.72} \\ \hline
\end{tabular}
\qquad
\begin{tabular}{|c|c | c|c|}\hline
\multicolumn{3}{|c|}{$\mu^+ \mu^+$ collision ($\sqrt{s}=2$ TeV)} \\ \hline
   & signal & BG  \\ \hline 
$N_{\rm ev}$ (without selection) &  55.8 & 282   \\ \hline
$N_{\rm ev}$ (with selection) & 2.64 & 0.00998    \\ \hline
$S$ & \multicolumn{2}{|c|}{4.93} \\ \hline
\end{tabular}
\caption{Left: Number of signal and BG events $N_{\rm ev}$ after selection, and corresponding discovery significance after selection for $\mu^+\mu^-$ collision with $\sqrt{s} = 3$ TeV where we choose  $m_{Z'} = 300$ GeV, $m_{\phi}=1500$ GeV, $g_X = 0.3$ as a benchmark point and integrated luminosity 100 fb$^{-1}$. Right: those for $\mu^+ \mu^+$ collision with $\sqrt{s} = 2$ TeV where we choose $m_{Z'} = 300$ GeV, $m_{\phi}=1200$ GeV, $g_X = 0.3$ as a benchmark point and integrated luminosity 100 fb$^{-1}$.  For number of events before selection we do not impose any cut and all $Z'/Z/W^\pm$ decay modes are included. }
\label{tab:Event}
\end{table}
\end{center}

We evaluate number of signal/BG events imposing selection of $3 \mu^+ \mu^-$ and $\mu^+\mu^+\mu^+\mu^+ \mu^- \mu^-$ for case (1) and (2) respectively with basic kinematical cut Eq.~\eqref{eq:Selection1} as $N_{\rm event} = L \sigma N_{Selected}/N_{\rm Generated}$ 
where $N_{\rm Select}$ is number of events after selection, $N_{\rm Generated}$ is number of originally generated events, $\sigma$ is a cross section for each process and $L$ is integrated luminosity.
We evaluate the discovery significance by
\begin{equation}
S = \sqrt{2 \left[ (N_S + N_{BG}  ) \ln \left( 1 + \frac{N_S}{N_{BG}} \right) - N_S\right]},
\end{equation}
where $N_S$ and $N_{BG}$ are respectively number of signal and BG events.
In Table~\ref{tab:Event}, we show number of signal/BG events before and after selection and discovery significance assuming integrated luminosity of 100 fb$^{-1}$ where we choose $m_{Z'} = 300$ GeV, $m_{\phi}=1500(1200)$ GeV and $g_X = 0.3$ as benchmark values for $\mu^-\mu^-(\mu^+\mu^+)$ collision cases; for number of events before selection we do not impose any cut and all $Z'/Z/W^\pm$ decay modes are included.
We find that sufficiently large significance is obtained for discovery with the benchmark point for $\mu^+ \mu^-$ collider while we get 
significance close to discovery level with the benchmark point for $\mu^+ \mu^+$ collider.
Furthermore we show the required integrated luminosity to realize $S=3(5)$ for both $\mu^+\mu^-$ and $\mu^+ \mu^+$ collider cases as functions of $m_{Z'}$ fixing $m_\phi = 1500$ GeV and $1200$ GeV, respectively, and $g'=0.3$.
We find that the required luminosity does not change much by changing $m_{Z'}$ value in the case of $\mu^+ \mu^-$ collider. 
This behavior is due to increase of cross section of $\mu^+ \mu^- \to Z' \phi$ process when we increase $Z'$ mass.
On the other hand the required luminosity increases as $m_{Z'}$ increases in the case of $\mu^+ \mu^+$ collider since $Z'$ fusion cross section decreases when $Z'$ mass increases for fixed $m_\phi$.
In the case of $\mu^+ \mu^-$ collider we can get discovery level significance with integrated luminosity below 50 fb$^{-1}$.
We can also obtain sufficiently large significance in the case of $\mu^+\mu^+$ collider when $Z'$ mass is light.

\begin{figure}[t]
\begin{center}
\includegraphics[width=70mm]{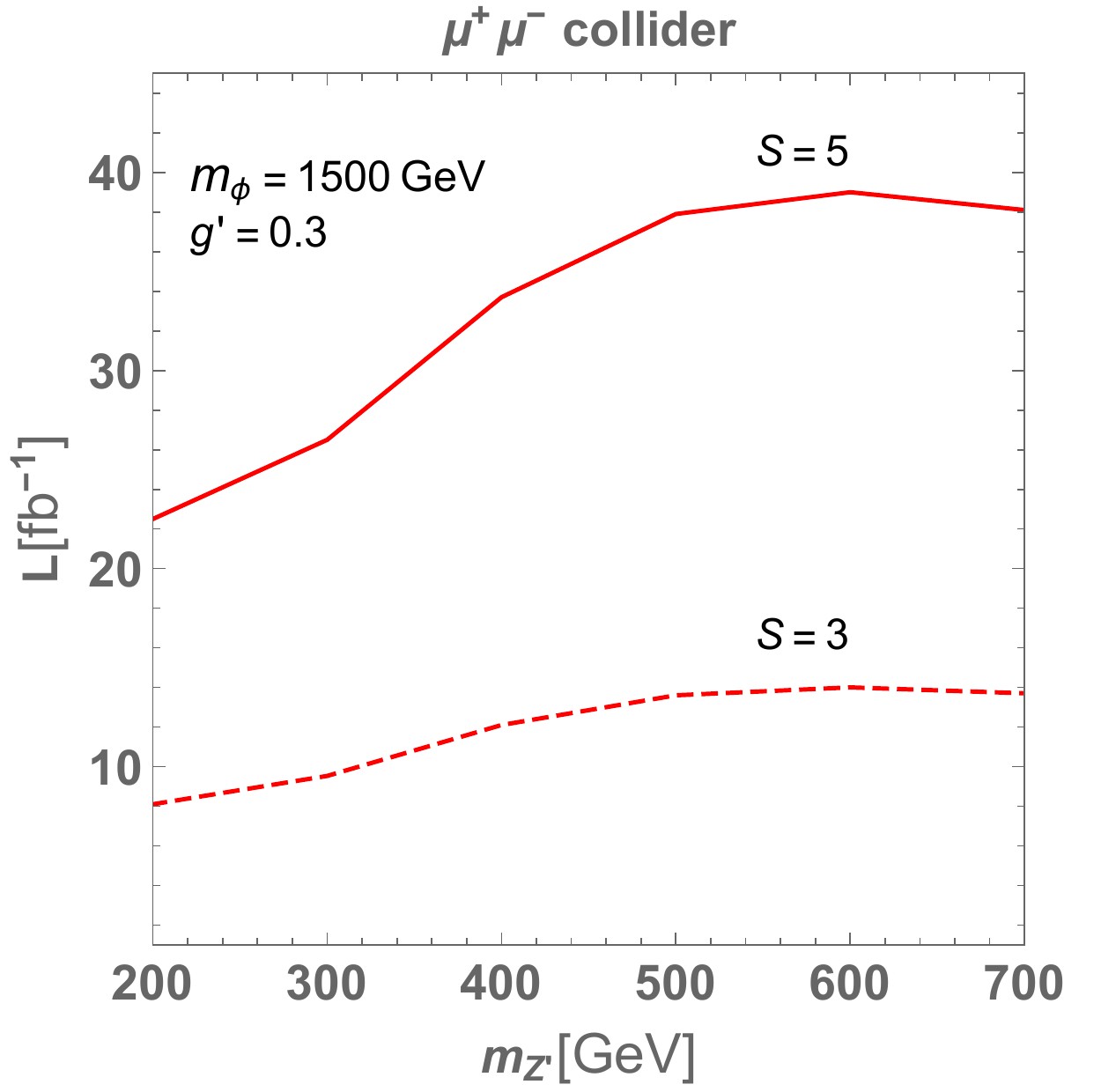} \quad
\includegraphics[width=70mm]{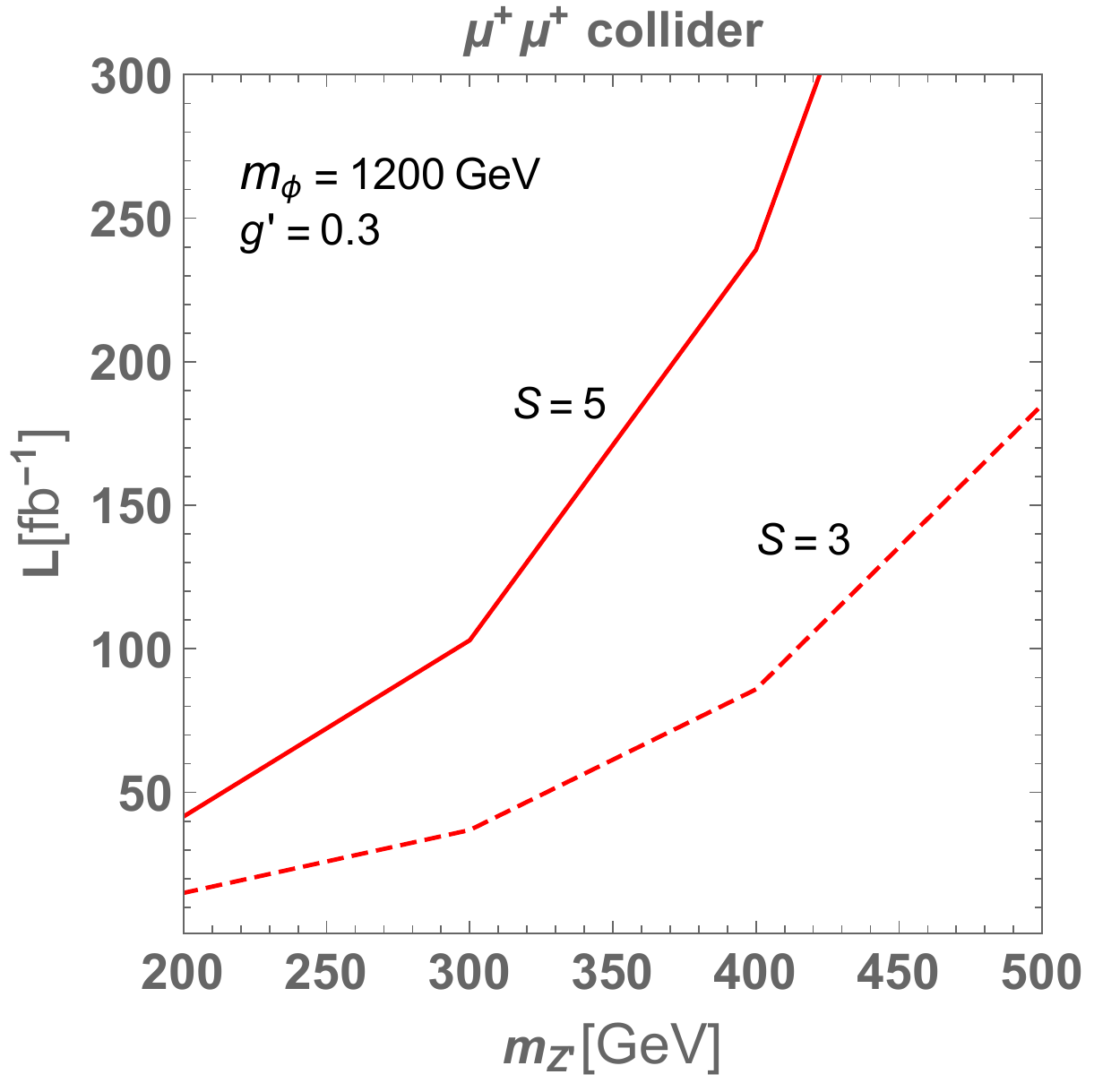}
\caption{The luminosity required to give significance of 3(5) for $\mu^+ \mu^-$ and $\mu^+ \mu^+$ collider cases as functions of $m_{Z'}$ where we fixed $m_\phi = 1500$ GeV and 1200 GeV, respectively, and $g'=0.3$.} 
  \label{fig:significance}
\end{center}\end{figure}

\section{Summary}

We have discussed signals of multi muon/anti-muon at $\mu^+ \mu^-$ and $\mu^+ \mu^+$ colliders 
that are induced from productions of new scalar boson $\phi$ with/without $Z'$  boson in a local $U(1)_{L_\mu - L_\tau}$ model.
The new scalar boson is associated with spontaneous breaking of the $U(1)_{L_\mu - L_\tau}$ gauge symmetry 
so that we have $\phi Z' Z'$ coupling.
The scalar boson dominantly decays into $Z'Z'$ mode and our $Z'$ boson decays into second and third generation leptons.

We focused on region of new gauge coupling and gauge boson mass that is favored to explain $b \to s \mu^+ \mu^-$ anomaly.
The cross sections of our signal process have been estimated applying benchmark parameter from the region and we have obtained  $\mathcal{O}(1)$ fb to $\mathcal{O}(10)$ fb values.
We have also carried out numerical simulation and found discovery significance where we can discover our signal with integrated luminosity less than 100 fb$^{-1}$ 
at our benchmark points.
Therefore muon colliders are quite suitable to test a spontaneously broken local $U(1)_{L_\mu - L_\tau}$ model.

\section*{Acknowledgments}
This work is supported by JSPS KAKENHI Grant Nos.~JP18K03651, JP18H01210, JP22K03622 and 
MEXT KAKENHI Grant No.~JP18H05543 (T.~S.), by the Fundamental Research Funds for the Central Universities (T.~N).

\end{document}